\documentstyle[aas2pp4]{article}       

\begin{document}

\title{The History of Cosmic Baryons: X-ray emission vs. Star Formation Rate}

\author{N. Menci}
\affil{Osservatorio Astronomico di Roma,
 via Osservatorio, 00040 Monteporzio, Italy}
\centerline{\&} 
\author{A. Cavaliere}
\affil{Astrofisica, Dipartimento Fisica, II Universit\`a di Roma,\\
via Ricerca Scientifica 1, 00133 Roma, Italy}

\begin{abstract}
Using  current models of structure and star formation based on 
hierarchical clustering, we connect the cosmic star formation rate
(SFR) to the X-ray emission properties of groups and clusters of
galaxies. We show that when the baryons cool down from a hot phase to
condense into stars with a SFR flat for $z\geq 2$, then such hot phase
yields deep X-ray source counts (at fluxes $F=10^{-15}$ erg/cm$^2$ s 
in the energy band 0.5-2 keV) about 5 times larger than in the
case of a SFR peaked at $z\approx 1.5$. We also discuss the effect of
the SFR on the shape and evolution of the X-ray $L-T$ relation.
\end{abstract}
\section{Introduction}
In current models of galaxy formation the SFR of the Universe is
tightly related to the amount of cold and hot gas ($m_c$ and $m_h$,
respectively) contained in the dark matter (DM) halos where the
galaxies are located. As a consequence, we expect the history of the SFR 
 to be related to the X-ray properties of groups and clusters
of galaxies, since these host most of the hot baryons emitting by
thermal bremsstrahlung luminosities $L\sim 10^{44}$ erg/s. In
particular, the process of ster formation is expected to affect the X-ray
properties of {\it small} galactic systems, since the ``stellar''
temperature $k T_* = E_*\, m_H/3 m_h\sim 0.2$ keV ($E_*$ being the
Supernova energy) which measures the thermal
energy provided by the Supernovae (SN) is close to the virial
temperature $kT \approx G\,M/10\,R \propto M^{2/3}$ of small galaxy
groups (DM masses $M\sim 10^{13}\,M_{\odot}$). Thus, in X-rays the
connection with the SFR should show up mainly at faint luminosities.

Since the behaviour of the cosmic SFR at $z>2$ is still plagued by
observational uncertainties in the O-IR bands, we explore here the
above connection with the X-ray data, and show that such observations
are within the reach of forthcoming X-ray observatories like AXAF.
These observations can discriminate between a declining SFR for $z>2$
(as initially proposed (Madau et al. 1996) and a flat SFR for $z>2$ (as
indicated by recent observations; see, e.g., Pettini 1997; Meurer et al. 1997).

The result may have relevant implications for all sides of the current
theories of galaxy formation, including the cosmogony of DM halos.  In
fact in these theories the SFR is strongly coupled to the dynamical
history of the DM halos, whose hierarchical growth is thought to 
proceed through merging and accretion events.

\section{Star Formation Models}

The baryonic processes inside dynamically evolving DM halos are
usually treated using semi-analytic models (SAM) in the framework of
hierarchical clustering (for a comprehensive account see Somerville \& 
Primack 1998).
According to the SAMs, the baryons gravitationally bound to a DM halo 
are assumed to be exchanged among three phases: the {\it cold} phase
produced by radiative cooling; the resulting {\it condensed} phase
into stars; the {\it hot} phase constituted by gas preheated by SN
explosions and further raised to the virial temperature of the
potential well. Besides, a fraction of the hot gas may be expelled by
SN winds; this depends on the key paremeter $\epsilon_o$, defining the
fraction of SN energy going into kinetic energy. The SAMs also include
coalescence of baryons following halo merging and the
luminosity-colour evolution due to rise and fall of successive star
generations. In the end, the statistics of the baryonic phases are
related to the statistics of the DM halos, usually
characterized in terms of their circular velocity $v_c$.

We developed an analytic version of the SAMs (based on the Extended
Press \& Schechter Theory, see Lacey \& Cole 1993), to describe the
above baryonic phases and the exchanges among them (we refer to
Menci \& Cavaliere 1999 for an extended description of the model and of its
results). The parametrizations we adopt are the same as in the  
SAMs. The star formation time scale and the SN feedback are related to
$v_c$ by scaling laws depending on a set of four global parameters
which, as described in Cole et al. 1994, are related to the detailed
parameter $\epsilon_o$. We focus on two particular sets of
parameters. The first (Model A) is characterized by a relatively large
value of $\epsilon_o=0.1$. With this, the SN winds are effective in
expelling hot gas from small halos.  Correspondingly, the resulting
SFR we find (see fig. 1) is peaked at $z\approx 1.5$ and then declines
considerably for larger $z$. The second set (Model B) is characterized
by a small $\epsilon_o=0.01$; in this case the SN are not effective in
expelling baryons from the halos, but they heat them to somewhat larger
temperatures compared with model A. 
\section{X-ray Observables Related to SFR}
The different conditions described by Models A and B lead to different
amounts and physical states for the hot baryons which are contained in the
DM potential wells and emit in X-rays. The X-ray luminosity of groups
or clusters may be written in the form (Cavaliere, Menci \& Tozzi 1998)
\begin{equation}
L\propto \rho^{1/2}(z)\,\bigg({m_h\over M}\bigg)^2\,I(M,z)\,G^2(M)\,T^2(M)~.
\end{equation}

Here the normalization is adjusted (as usually) so as to match the
height of the observed {\it local} $L-T$ relation at $T=4$ keV;
$\rho(z)$ is the DM density of the halo; the shape factor $I$
describes the internal ICM distribution. In addition, $G$ denotes the
density jump across the shock induced by the infalling gas, averaged
over the merging histories, that is, over the probability to merge 
clumps $M'$ onto $M$; $G$ is governed by $T/T'$, where $T'$ is the
temperature of infalling gas. The overall effect of $G^2$ is to
depress the $L-T$ relation at the scale of groups, bending it down
from the self-similar behaviour $L\propto T^2$, which
applies to very rich clusters Ponman et al. (1996). 

The different amounts of hot baryons retained in small halos in Model
A and Model B affect the $z$-evolution of the $L-T$ relation (see
fig. 2); the considerable {\it increase} with $z$ at the faint end of
$L$ found in Model B reflects the larger amount of emitting baryons in the
smaller halos, which dominate the statistics at $z> 1$. In
addition, the larger amount of expelled diffuse baryons marking Model
A, together with the corresponding lower SN-induced temperature,
make the shocks more effective in this case. Correspondingly, the
factor $G^2$ in eq. 1 is more effective in bending down the $L-T$
relation from its self-similar shape.
 
Such effects of different SFR on the X-ray emission at the groups scale 
interestingly affect the deep source counts, as shown in fig. 3.
 Note that at fluxes $F\approx 10^{-15}$ erg/s cm$^2$ 
(reachable, e.g., by AXAF) a flat SFR implies counts 
larger by a factor $\gtrsim 5$ with respect to a SFR declining for $z>2$.

\newpage
\section*{FIGURE CAPTIONS}
\figcaption[]{ 
The SF histories in Model A (left panel) and model B
 (right panel). 
Critical Universe with $H_o=50$ km/s Mpc and tilted CDM spectrum of 
perturbations.
\label{fig1}}

\figcaption[]{ The $L-T$ correlation at $z=0$ (solid line) and $z=1$ 
(dashed) in Model A (left) and Model B (right). 
The data are taken from Ponman et al. (1996) and Markevitch (1998).
\label{fig2}}

\figcaption[]{ The X-ray source counts in the energy band 0.5-2 keV for the 
SFRs derived from Model A (solid line) and from Model B (dashed line). 
The assumed cosmological/cosmogonical parameters are as given for 
fig. 1. The shaded region corresponds to the 
ROSAT cluster counts observed by Rosati et a. (1998).
\label{fig3}}


\begin{references}
%
\addcontentsline{toc}{section}{References}

\reference{}Allen, S.M, \& Fabian, A.C. (1998), Monthly
Not. Roy. Astron. Soc., 297, L57

\reference{}Cavaliere, A., Menci, N., Tozzi, P. (1998), in press [astro-ph9810498]

\reference{}Cole, S., Aragon-Salamanca, A., Frenk, C.S., Navarro, J.F., \& 
 Zepf, S.E. (1994), Monthly Not. Roy. Astron. Soc., {\bf 271}, 781

\reference{}Lacey, C., \& Cole, S. (1993),  Monthly Not. Roy. Astron. Soc.,
{\bf 262}, 627

\reference{}Madau, P., Ferguson, H.C., Dickinson, M.E., Giavalisco, M., 
Steidel, C.C., \& Fruchter, A. (1996), Monthly Not. Roy. Astron. Soc., 
{\bf 283}, 1388

\reference{}Markevitch, M. (1998), Astrophys. Journ., {\bf 504}, 27

\reference{}Menci, N., \& Cavaliere, A. (1999), preprint 

\reference{}Meurer, G. Heckman, T.M., 
Lehnert, M.D., Leitherer,  C., \& Lowenthal, J. (1997), Astron. Journ., 
{\bf 111}, 54

\reference{} Pettini, M. et al. (1997), {\it The Spectra of Star Forming 
Galaxies at High Redshift}, in {\it The Ultraviolet Universe at 
Low and High Redshift: Probing the Process of Galaxy Evolution}, 
ed. Waller W. et al., AIP Conf. Proc. 408, 279

\reference{}Ponman, T.J., Bourner, P.D.J., Ebeling, H., B\"ohringer, 
H. 1996, Monthly Not. Roy. Astron Soc., {\bf 283}, 690

\reference{} Rosati, P., Della Ceca, R., Norman, C., \& Giacconi, R. 
(1998),  Astrophys. Journ., {\bf 492}, 21

\reference{}Sommerville, R.S., \& Primack, J.R. (1998), 
preprint [astro-ph/9802269]  


\end{references}
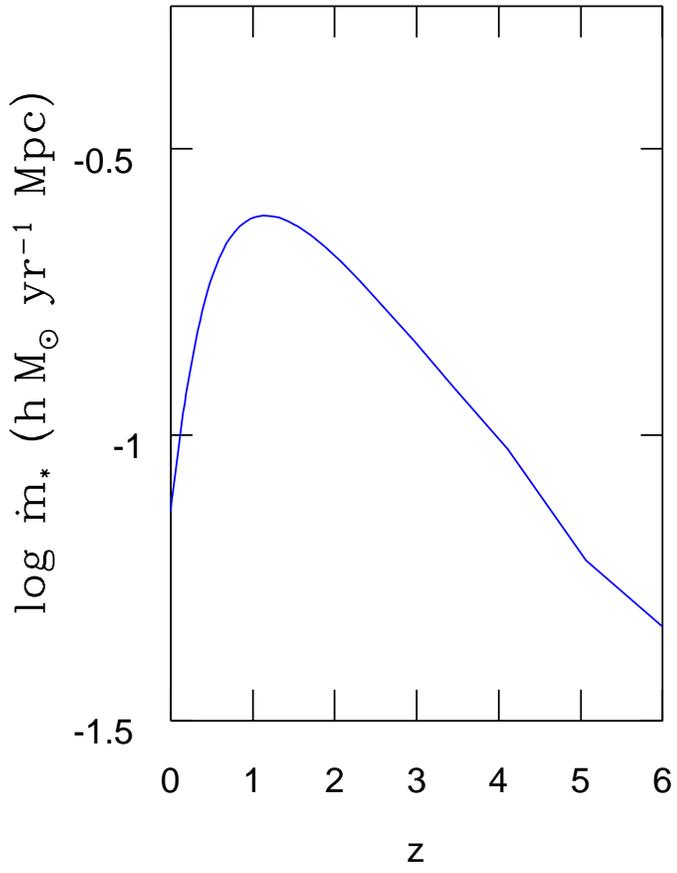
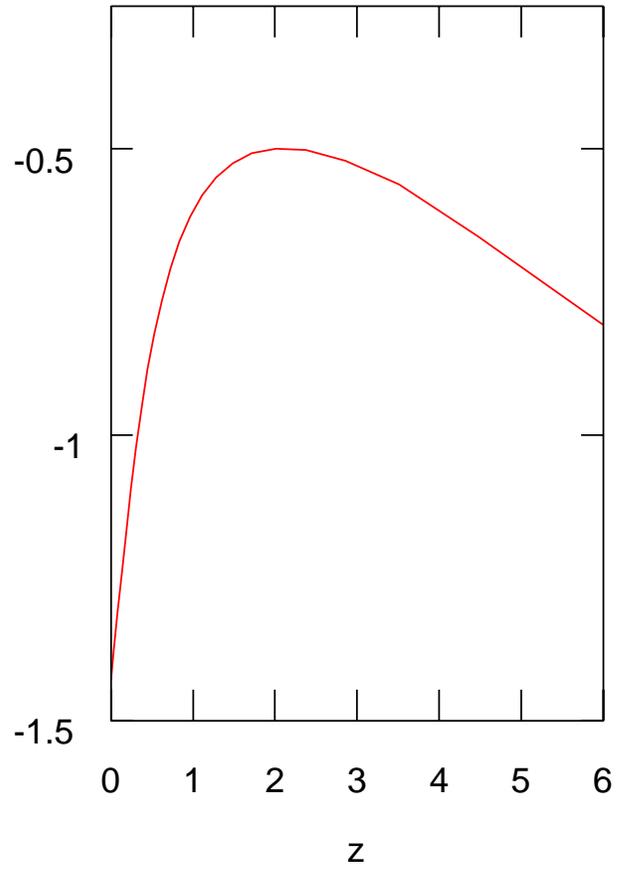
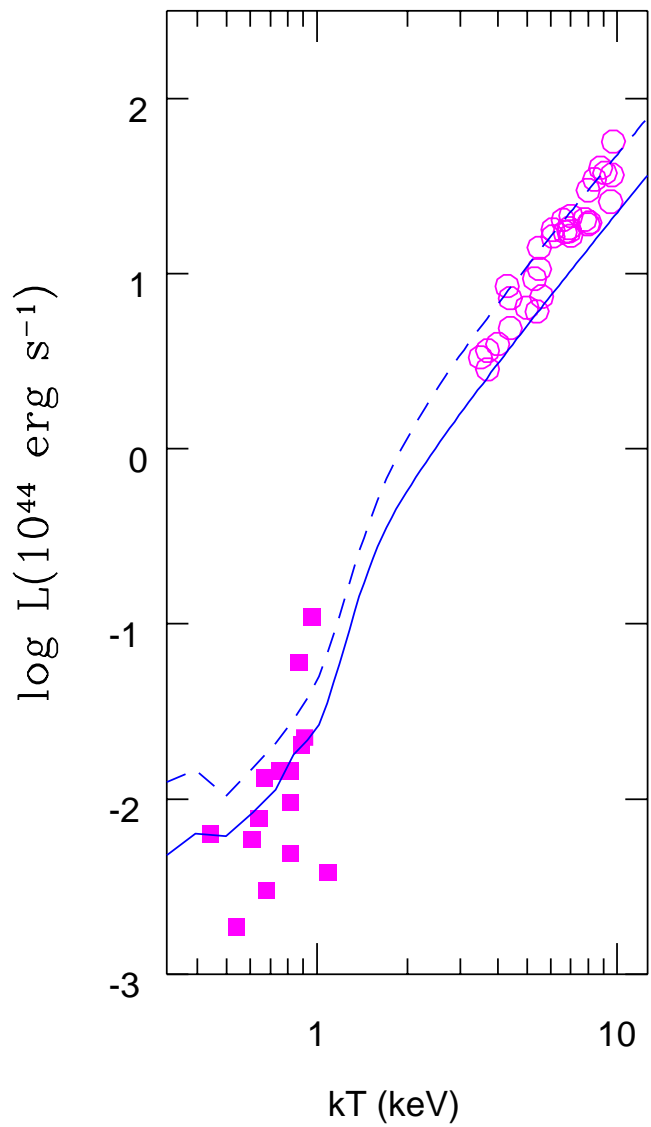
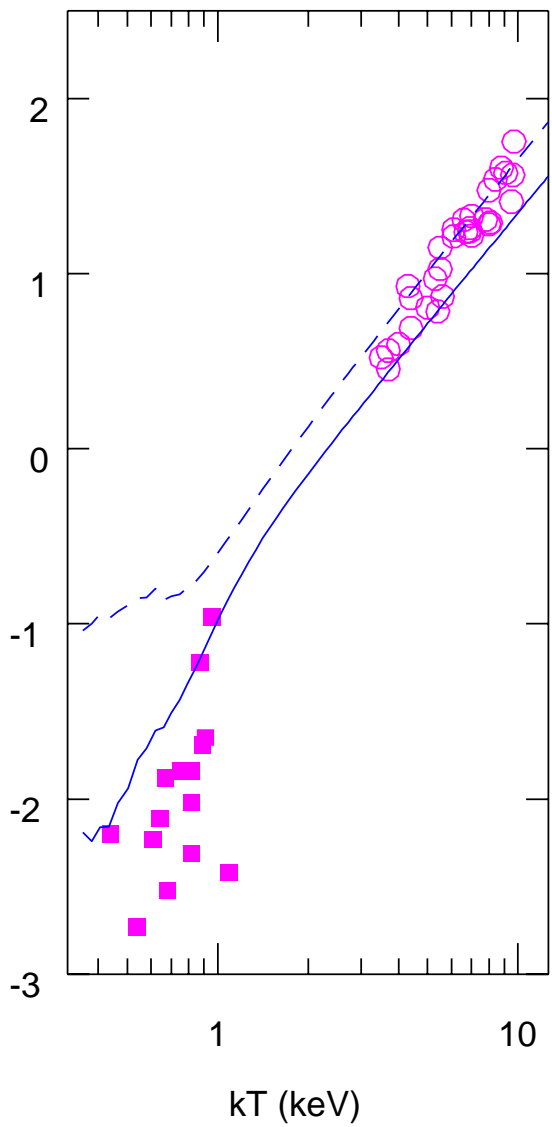
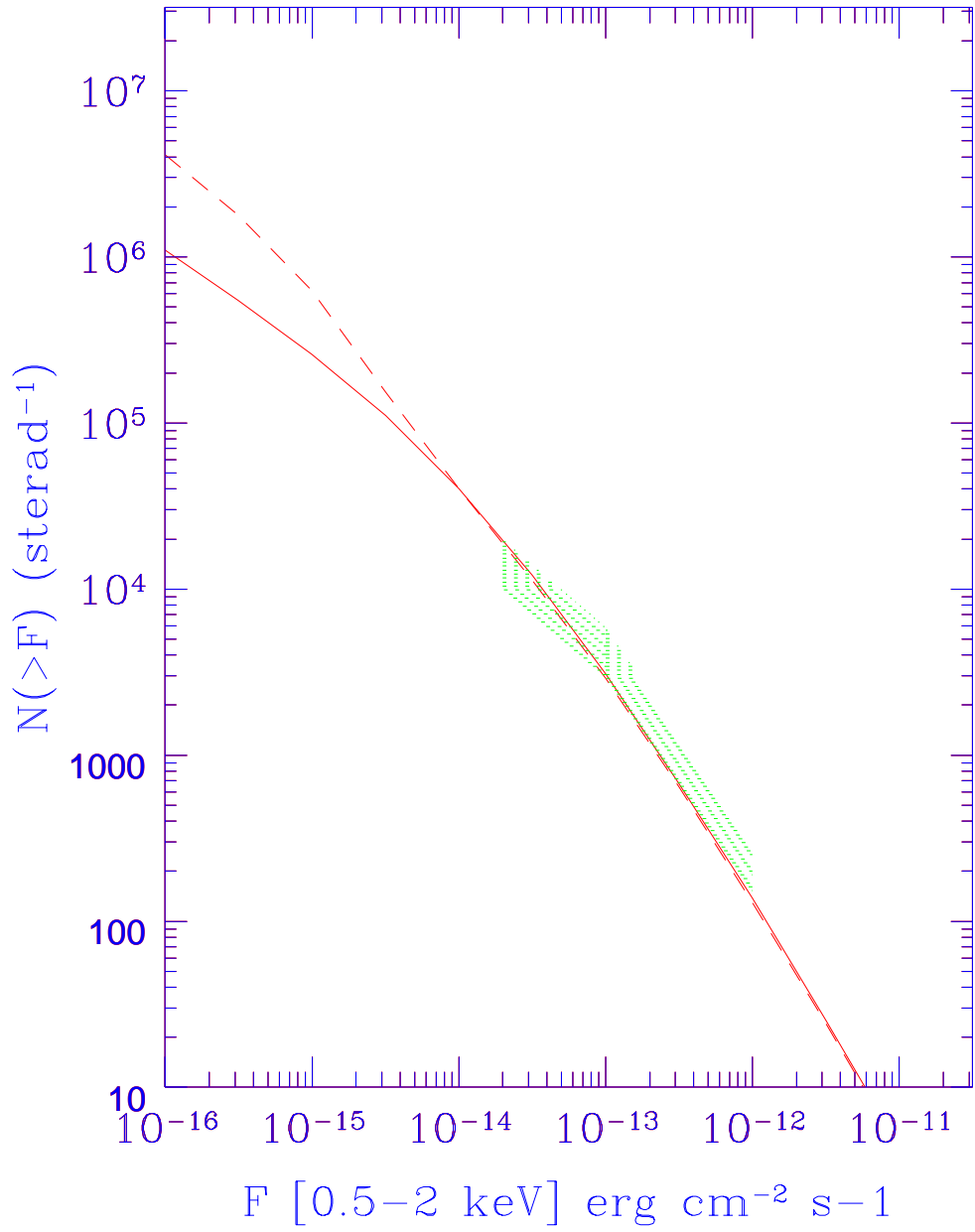
\end{document}